\documentclass{IEEEtran}

\usepackage{amsmath,amssymb,amsthm}

\usepackage{tikz}
\usetikzlibrary{external}

\allowdisplaybreaks

\newcommand{\F}{\mathbb{F}}

\newcommand{\Z}{\mathbb{Z}}

\newcommand{\ga}{\alpha}

\newcommand{\gam}{\gamma}

\newcommand{\gd}{\delta}

\newcommand{\gw}{\omega}
\renewcommand{\phi}{\varphi}
\newcommand{\bL}{\bar\Lambda}
\newcommand{\gL}{\Lambda}

\newcommand{\calL}{\mathcal{L}}

\newtheorem{thm}{Theorem}
\newtheorem{lem}[thm]{Lemma}
\newtheorem{prop}[thm]{Proposition}

\newcommand{\set}[1]{\{#1\}}

\newcommand{\fl}[1]{\lfloor{#1}\rfloor}

\newcommand{\yb}{\bar{y}}
\newcommand{\y}{\mathsf{y}}
\newcommand{\x}{\mathsf{x}}
\newcommand{\ev}{\mathrm{ev}}
\newcommand{\wt}{\mathrm{wt}}
\newcommand{\xdeg}{\deg_x}
\newcommand{\ydeg}{\deg_y}
\newcommand{\ybdeg}{\deg_{\yb}}
\newcommand{\zdeg}{\deg_z}

\newcommand{\dLO}{d_\mathrm{LO}}
\newcommand{\Rbar}{\bar{R}}
\newcommand{\lbar}{s}

\DeclareMathOperator{\LT}{lt}
\DeclareMathOperator{\LM}{lm}
\DeclareMathOperator{\LC}{lc}

\begin{document}

\title{Unique Decoding of General AG Codes}

\author{Kwankyu Lee, Maria Bras-Amor\'os, and Michael E.~O'Sullivan
\thanks{K.~Lee is with the Department of Mathematics, Chosun University, Gwangju 501-759, Korea (e-mail: kwankyu@chosun.ac.kr). He was supported by Basic Science Research Program through the National Research Foundation of Korea(NRF) funded by the Ministry of Education, Science and Technology(2009-0064770) and also by research fund from Chosun University, 2008.}%
\thanks{M.~Bras-Amor\'os is with the Department of Computer Engineering and Mathematics, Universitat Rovira i Virgili, Tarragona 43007, Catalonia, Spain (e-mail: maria.bras@urv.cat). She was supported by the Spanish Government through the projects TIN2009-11689 ``RIPUP" and CSD2007-00004 ``ARES".}%
\thanks{M.~E.~O'Sullivan is with the Department of Mathematics and Statistics, San Diego State University, San Diego, CA 92182-7720, USA (e-mail: mosulliv@math.sdsu.edu). He was supported by the National Science Foundation under Grant No. CCF-0916492.}%
}

\maketitle

\begin{abstract}
A unique decoding algorithm for general AG codes, namely multipoint evaluation codes on algebraic curves, is presented. It is a natural generalization of the previous decoding algorithm which was only for one-point AG codes. As such, it retains the same advantages of fast speed and regular structure with the previous algorithm. Compared with other known decoding algorithms for general AG codes, it is much simpler in its description and implementation.
\end{abstract}

\IEEEpeerreviewmaketitle


\section{Introduction}

Goppa \cite{goppa1981} was the first to define linear error-correcting codes on algebraic curves. For a divisor $G$ whose support is disjoint from a set of rational points on the curve, divisor $D$ being the sum of those rational points, he defined the evaluation code $C_\calL(D,G)$ and the differential code $C_\Omega(D,G)$, the latter being the dual of the former. In the subsequent vast research works on Goppa's codes, now called AG codes, the focus was often on the dual of the evaluation code, that is, the differential code. The reason seems to be nothing else but the first successful decoding algorithm for AG code \cite{justesen1989} was for the dual of the evaluation codes. Thus a lot of effort was put into finding curves with many rational points and thereon to construct differential codes with good parameters. To estimate the minimum distance of the codes, various lower bounds have been developed. For much the same reason, so-called one-point codes that assume $G=mQ$ for some positive integer $m$ and a rational point $Q$ are considered most often in the literature. These one-point differential codes can be decoded efficiently by the syndrome-based Berlekamp-Massey-Sakata algorithm with the Feng-Rao majority voting \cite{osullivan1995}.  

Guruswami and Sudan's list decoding \cite{guruswami1999} provided a fresh point of view that brought the evaluation codes back to the center. Using interpolation, they showed that evaluation codes can be decoded successfully beyond the capacity of the previous decoding algorithms for differential codes. Following this way of approaching the decoding problem of AG codes, the authors \cite{kwankyu2011} reinterpreted Duursma's idea of the majority voting \cite{duursma1993} in the context of the interpolation decoding, and introduced a unique decoding algorithm for one-point evaluation codes on Miura-Kamiya plane curves. The result was a combination of nice features of the interpolation-based list decoding and the performance of the classical syndrome decoding with the majority voting scheme.  Shortly thereafter, Geil et al.~\cite{geil2012b} generalized the result for arbitrary one-point AG codes and for list decoding. The goal of this paper is to note that the basic idea of \cite{kwankyu2011} is more widely applicable, and present an interpolation-based unique decoding algorithm for general evaluation AG codes. By general evaluation AG codes, we mean the evaluation codes $C_\calL(D,G)$ with an arbitrary divisor $G$, with the premise that there exists a rational point $Q$ not in the support of $D$. These codes are often called multipoint evaluation codes. Prominent examples would be the two-point codes on maximal curves such as Hermitian, Suzuki, and Klein curves.

We find that the impact of the interpolation-based list decoding has already made Beelen and H\o holdt \cite{beelen2007} to construct a unique decoding algorithm that is very similar to ours. Their algorithm also adopts an iterative method using majority voting to find the interpolation polynomial that gives the corrected codeword. The major difference of our algorithm is that we do not need differentials to construct the algorithm and use Lagrange interpolation instead of syndromes computed from the received vector, and thus directly compute the coefficients, corresponding to the sent message, by majority voting. Thus our algorithm is much simpler to present and more streamlined to implement and deploy in practice. Fujisawa and Sakata \cite{fujisawa2011} also presented a fast decoding algorithm for multipoint general AG codes using a variant of the classical Berlekamp-Massey-Sakata algorithm, but only to correct errors short of the Goppa bound. Their method, originally due to Drake and Matthews \cite{drake2010}, is to embed the multipoint code isometrically into a one-point code.  

The core ideas of the present work that we add to \cite{kwankyu2011} are all contained in the preliminary materials in Section \ref{kkmcd}. For general facts and notations for algebraic curves and functions fields, we refer to \cite{stichtenoth2009}. Once the stage set, we describe in Section \ref{sec_dwkqd} the decoding algorithm in a parallel fashion to \cite{kwankyu2011}. In Section \ref{iqwkdq}, several examples and experimental results are provided. In the final Section, we conclude with some remarks.

\section{Preliminaries}\label{kkmcd}

Let $X$ be a smooth geometrically irreducible projective curve defined over a finite field $\F$. Let $P_1,P_2,\dots,P_n$ and $Q$ be distinct rational points on $X$, and define $D=P_1+P_2+\dots+P_n$. Let $G$ be an arbitrary divisor on $X$, whose support is disjoint from that of $D$, but allowed to include $Q$.

Let $\F(X)$ be the function field of $X$ over $\F$. Let
\[
	R=\bigcup_{s=0}^\infty\calL(sQ)\subset\F(X)
\]
be the ring of all functions on $X$ which have no poles other than $Q$. For $f\in R$, let $\rho(f)=-v_Q(f)$. The Weierstrass semigroup at $Q$ is then
\[
	\gL =\set{\rho(f)\mid f\in R}.
\]
It is well-known that $\gL$ is a numerical semigroup whose number of gaps is the genus $g$ of $X$. Let $\gamma$ be the smallest positive integer in $\gL$, and let $\rho(x)=\gamma$ with some $x\in R$. For each $0\le i<\gamma$, let $a_i$ be the smallest integer such that $a_i\equiv i\pmod{\gam}$ and $\rho(y_i)=a_i$ for some $y_i\in R$. Then, using the properties of $\rho:R\to\Z_{\ge 0}$ inherited from the valuation $v_Q$, we can show that $\set{y_0,y_1,\dots,y_{\gam-1}}$ forms a basis of $R$ as a free module of rank $\gam$ over $\F[x]$. Hence $\set{x^k y_i\mid k\ge 0, 0\le i<\gam}$ is a vector space basis of $R$ over $\F$, and will be called the monomials of $R$. The set $\set{a_i\mid 0\le i<\gam}$ is usually referred to as the Ap\'ery set of $\gL$. 

Now let	
\[
	\Rbar=\bigcup_{s=-\infty}^\infty\calL(sQ+G)\subset\F(X),
\]
which is clearly a module over $R$. For $f\in \Rbar$, let $\gd(f)$ denote the smallest integer $s$ such that $f\in\calL(sQ+G)$. Note that simply $\gd(f)=-v_Q(f)-v_Q(G)$. Thus the map $\gd:\Rbar\to\Z$ satisfies the following properties:
\begin{itemize}
\item[(1)] $\gd(f)\ge -|G|$ for $f\in \Rbar$, where $|G|=\deg(G)$.
\item[(2)] $\gd(fg)=\rho(f)+\gd(g)$ for $f\in R$, $g\in \Rbar$.
\item[(3)] $\gd(f+g)\ge\max\set{\gd(f),\gd(g)}$ for $f,g\in \Rbar$. The equality holds if $\gd(f)\neq\gd(g)$.
\item[(4)] If $\gd(f)=\gd(g)$, then there is a unique $c\in\F$ such that $\gd(f)>\gd(f-cg)$. 
\end{itemize}
Let 
\[
	\bL=\set{\gd(f)\mid f\in \Rbar}=\set{\lbar_0,\lbar_1,\lbar_2,\dots}.
\]
Then $\gL+\bL=\bL$, and hence $\bL$ contains all large enough integers. Therefore for each $0\le i<\gamma$, there exists the smallest integer $b_i$ such that $b_i\equiv i\pmod{\gam}$ and $\gd(\yb_i)=b_i$ for some $\yb_i\in \Rbar$. Then using the properties of $\gd$, we easily see that $\set{\yb_i\mid 0\le i<\gamma}$ forms a basis of $\Rbar$ as a free module of rank $\gam $ over $\F[x]$. For $s\in \bL$, if $i=s\mod\gam$ and $k=(s-b_i)/\gam\ge 0$, define $\phi_s=x^k \yb_i$. Note that $\gd(\phi_s)=s$. Thus $\set{\phi_s\mid s\in\bL}=\set{x^k\yb_i\mid k\ge 0, 0\le i<\gam}$ is a basis of $\Rbar$ over $\F$, and will be called the monomials of $\Rbar$. 

Let us consider the $R$-module
\[
	Rz\oplus \Rbar=\set{fz+g\mid f\in R, g\in \Rbar},
\]
where $z$ is a variable. Note that it is also a free $\F[x]$-module of rank $2\gamma$ with free basis
\[
	K=\set{y_iz,\yb_i\mid 0\le i<\gamma}.
\]
Thus every element in $Rz\oplus \Rbar$ can be written as a unique $\F$-linear combination of the monomials in
\[
	\Omega=\set{x^ky_iz,x^k\yb_i \mid k\ge 0, 0\le i<\gam}.
\]
For the monomials, we will use the notations
\[
	\begin{aligned}
	\xdeg(x^ky_iz)&=k, & \ydeg(x^ky_iz)&=i, & \zdeg(x^ky_iz)&=1,\\
	\xdeg(x^k\yb_i)&=k, &\ybdeg(x^k\yb_i)&=i, & \zdeg(x^k\yb)&=0.\\
	\end{aligned}
\]

We now briefly review the Gr\"obner basis theory on $Rz\oplus \Rbar$, regarded as a free module of rank $2\gam$ over $\F[x]$. First we define monomial order $>_s$. For an integer $s$, the weighted degree of a polynomial $fz+g\in Rz\oplus \Rbar$ is defined as
\[
	\gd_s(fz+g)=\max\set{\rho(f)+s,\gd(g)}.
\]
In particular, for monomials, we have 
\[
	\begin{aligned}
	\gd_s(x^ky_iz)&=\gamma k+a_i+s,\\
	\gd_s(x^k\yb_i)=\gd(x^k\yb_i)&=\gam k+b_i.
	\end{aligned}
\]
Then $\gd_s$ induces the weighted degree order $>_s$ on $\Omega$, where we break ties by declaring the monomial with $z$ precedes the other without $z$. For $f\in Rz\oplus \Rbar$, the notations $\LT_s(f)$, $\LM_s(f)$, and $\LC_s(f)$ are used to denote respectively the leading term, the leading monomial, and the leading coefficient, with respect to $>_s$. If $f\in \Rbar$, we may omit the superfluous $s$ from these notations. Finally there is a simple criterion to recognize a Gr\"obner basis of an $\F[x]$-submodule of $Rz\oplus \Rbar$. 

\begin{prop}\label{xnskw}
Let $S$ be a submodule of $Rz\oplus \Rbar$, and $B$ generate $S$ over $\F[x]$. If elements of $B$ have leading terms with respect to $>_s$ that are $\F[x]$-multiples of distinct elements of $K$, then $B$ is a Gr\"obner basis of $S$ with respect to $>_s$. If this is the case, $B$ is also a free basis of $S$.    
\end{prop}

For more discussion on Proposition \ref{xnskw} and on the general theory of Gr\"obner bases, we refer to \cite{cox2005}.

The evaluation map 
\[
	\ev:\Rbar\to\F^n,\quad\phi\mapsto(\phi(P_1),\phi(P_2),\dots,\phi(P_n))
\]
is linear over $\F$. Thus the AG code 
\[
	C=C_\calL(D,G)=\ev(\calL(G))
\]
is a linear code of length $n$ over $\F$. Let us assume $|G|<n$ so that the functions in $\calL(G)$ correspond one-to-one with the codewords in $C$ under $\ev$. Note that $\set{\phi_s\mid s\in\bL,s\le 0}$ is a basis of $\calL(G)$ as a vector space over $\F$. Hence the dimension of $C$ is $k=|\set{s\in\bL\mid s\le 0}|$. So $\set{s\in\bL\mid s\le 0}=\set{s_0,s_1,\dots,s_{k-1}}$. We will also assume the nonsystematic encoding by evaluation. Thus a message $\gw=(\gw_{s_0},\gw_{s_1},\dots,\gw_{s_{k-1}})\in\F^k$ is encoded to the codeword $\ev(\mu)\in C$ where 
\[
	\mu=\sum_{i=0}^{k-1}\gw_{s_i}\phi_{s_i}\in\calL(G).
\]

Note that the map $\ev$ is surjective onto $\F^n$. Indeed by the Riemann-Roch theorem, we see that $\ev(\calL(sQ+G))=\F^n$ for $s\ge n-|G|+2g-1$. Let $h_i\in \Rbar$ be such that $\ev(h_i)$ is the $i$th element of the standard basis of $\F^n$. Let $J$ be the kernel of $\ev$. Note that $J$ is a submodule of $\Rbar$ over $R$, and also over $\F[x]$. Let $\set{\eta_i\mid 0\le i<\gam}$ be a Gr\"obner basis of $J$ over $\F[x]$ such that $\ybdeg(\LT(\eta_i))=i$.

\begin{prop}\label{camdzs}
We have
\[
	\sum_{0\le i<\gam}\xdeg(\LT(\eta_i))=\dim_\F \Rbar/J=n.
\] 
\end{prop}

\begin{IEEEproof}
The first equality is a standard result of the Gr\"obner basis theory. To see the second equality, note that for all large enough $s$, 
\[
	\dim_\F \Rbar/J=\dim_\F\calL(sQ+G)/\calL(sQ+G-\sum_{i=1}^n P_i)=n.
\]
\end{IEEEproof}

Now let $v\in \F^n$ be the received vector. Suppose $c\in C$ is such that $v=c+e$, where $c=\ev(\mu)$ for a unique
\[
	\mu=\sum_{s\in\bar{\gL},s\le 0}\gw_s\phi_s\in\calL(G).
\]
The goal of a decoding algorithm is to recover $\mu$, and also $c$ if necessary, from $v$. We consider the interpolation module
\[
	I_v=\set{fz+g\in Rz\oplus \Rbar\mid f(P_i)v_i+g(P_i)=0, 1\le i\le n}.
\]
Using the Gr\"obner basis theory, we will extract $\mu$ from $I_v$. 

Let 
\[
	h_v=\sum_{i=1}^nv_i h_i
\]
so that $\ev(h_v)=v$. Then $I_v=R(z-h_v)+J$. Hence by the criterion in Proposition \ref{xnskw}, the set
\begin{equation}\label{cjsjsw}
	\set{y_i(z-h_v),\eta_i\mid 0\le i<\gam}
\end{equation}
is a Gr\"obner basis of $I_v$ with respect to $>_{\gd(h_v)}$.

The ideal of the error vector $e$
\[
	J_e=\bigcup_{s=0}^\infty\calL(sQ-\sum_{e_i\neq 0}P_i)\subset R
\]
is also a submodule of $R$ over $\F[x]$, and has a Gr\"obner basis $\set{\epsilon_i\mid 0\le i<\gam}$ with respect to $>_s$ such that $\ydeg(\LT(\epsilon_i))=i$. We prove the following by the same argument as before.

\begin{prop}\label{mdcad}
We have
\[
	\sum_{0\le i<\gam}\xdeg(\LT(\epsilon_i))=\dim_\F R/J_e=\wt(e).
\]
\end{prop}

\section{Decoding Algorithm}\label{sec_dwkqd}

Notice that this section is adapted from the corresponding section in \cite{kwankyu2011} for the present general setup, with some changes in notations. Some minor errors are also corrected.

\subsection{Theory}

The basic idea of our decoding algorithm is to iteratively compute the coefficients $\gw_s$ of the function $\mu$. For $s>0$, define $v^{(s)}=v$, $c^{(s)}=c$, and $\mu^{(s)}=\mu$. For $s\in\bL, s\le 0$, define
\[
\begin{aligned}
	\mu^{(s-1)}&=\mu^{(s)}-\gw_s\phi_s,\\
	c^{(s-1)}&=c^{(s)}-\ev(\gw_s\phi_s),\\
	v^{(s-1)}&=v^{(s)}-\ev(\gw_s\phi_s),
\end{aligned}
\]
and for $s\not\in\bL, s\le 0$, let $v^{(s-1)}=v^{(s)}$, $c^{(s-1)}=c^{(s)}$, and $\mu^{(s-1)}=\mu^{(s)}$. Note that 
\[
	\mu^{(s)}\in\calL(sQ+G),\quad
	c^{(s)}=\ev(\mu^{(s)}),\quad
	v^{(s)}=c^{(s)}+e	
\]
for all $s$. Let $B^{(s)}=\set{g_i^{(s)},f_i^{(s)}\mid 0\le i<\gam}$,
\[
	\begin{aligned}
	g_i^{(s)}&=\sum_{0\le j<\gam}c_{i,j}y_jz+\sum_{0\le j<\gam}d_{i,j}\yb_j\\
	f_i^{(s)}&=\sum_{0\le j<\gam}a_{i,j}y_jz+\sum_{0\le j<\gam}b_{i,j}\yb_j
	\end{aligned}
\]
be a Gr\"obner basis of $I_{v^{(s)}}$ with respect to $>_s$ satisfying the criterion $\LT_s(g_i^{(s)})=\LT(d_{i,i}\yb_i)$ and $\LT_s(f_i^{(s)})=\LT_s(a_{i,i}y_iz)$, where $a_{i,j},b_{i,j},c_{i,j},d_{i,j}\in\F[x]$, for which we suppress the necessary superscript {\scriptsize$(s)$} for legibility.

\begin{lem}\label{lem_jjqqe}
We have
\[
	\sum_{0\le i<\gam}\deg(a_{i,i})+\sum_{0\le i<\gam}\deg(d_{i,i})=n.
\]
\end{lem}

\begin{IEEEproof}
As $B^{(s)}$ is a Gr\"obner basis of $I_{v^{(s)}}$,
\[
	\sum_{0\le i<\gam}\deg(a_{i,i})+\sum_{0\le i<\gam}\deg(d_{i,i})=\dim_{\F}(Rz\oplus \Rbar)/I_{v^{(s)}}.
\]
Recall that $I_{v^{(s)}}=R(z-h_{v^{(s)}})+J$. Hence $\dim_{\F}(Rz\oplus \Rbar)/I_{v^{(s)}}=\dim_\F \Rbar/J=n$.
\end{IEEEproof}

\begin{lem}
For $0\le i<\gam$, we have $\rho(a_{i,i}y_i)\le\rho(\epsilon_i)$, that is, $\deg(a_{i,i})\le \xdeg(\LT(\epsilon_i))$.
\end{lem}

\begin{IEEEproof}
Since $J_e(z-\mu^{(s)})\subset I_{v^{(s)}}$, we have $\epsilon_i(z-\mu^{(s)})\in I_{v^{(s)}}$. Note that $\LT_s(\epsilon_i(z-\mu^{(s)}))=\LT_s(\epsilon_iz)$. As $B^{(s)}$ is a Gr\"obner basis of $I_{v^{(s)}}$, the leading term $\LT_s(\epsilon_iz)$ must be an $\F[x]$-multiple of $\LT_s(f_i^{(s)})$. Therefore $\gd_s(a_{i,i}y_iz)\le\gd_s(\epsilon_iz)$ so that $\rho(a_{i,i}y_i)\le\rho(\epsilon_i)$.
\end{IEEEproof}

\begin{lem}\label{xnndd}
For $0\le i<\gam$, we have $\gd(d_{i,i}\yb_{i})\le \gd(\eta_{i})$, that is $\deg(d_{i,i})\le \xdeg(\LT(\eta_i))$.
\end{lem}

\begin{IEEEproof}
As $B^{(s)}$ is a Gr\"obner basis of $I_{v^{(s)}}$ and $J\subset I_{v^{(s)}}$, it follows that $\LT(\eta_{i})$ is an $\F[x]$-multiple of $\LT_s(g_{i}^{(s)})$. Hence $\gd(d_{i,i}\yb_i)\le \gd(\eta_{i})$.
\end{IEEEproof}

Now let $w$ be an element of $\F$. For each $0\le i<\gam$, let
\begin{equation*}
	\hat{g}_{i}=g_{i}^{(s)}(z+w\phi_s),\quad
	\hat{f}_i=f_i^{(s)}(z+w\phi_s)
\end{equation*}
where the parentheses denote substitution of the variable $z$. The automorphism of the module $Rz\oplus \Rbar$ induced by the substitution $z\mapsto z+w\phi_s$ preserves leading terms with respect to $>_s$. Therefore the set $\hat{B}=\set{\hat{g}_i,\hat{f}_i\mid 0\le i<\gam}$ is a Gr\"obner basis of 
\[
	\tilde{I}=\set{f(z+w\phi_s)\mid f\in I_{v^{(s)}}}
\]
with respect to $>_s$. However, with respect to $>_{s-1}$, $\hat{B}$ may not be a Gr\"obner basis of $\tilde{I}$. The following procedure modifies $\hat{B}$ to obtain a Gr\"obner basis of $\tilde{I}$ with respect to $>_{s-1}$.

For each $0\le i<\gam$, there are unique integers $0\le i'<\gam$ and $k_i$ satisfying
\begin{equation}\label{vnmskq}
	\rho(a_{i,i}y_i)+s=\gam k_i+b_{i'}
\end{equation}
such that $\rho(a_{i,i}y_i)+s\in\bL$ if and only if $k_i\ge 0$. Let
\begin{equation}\label{jcmkwd}
	c_i=\deg(d_{i',i'})-k_i,\quad \bar{c}_i=\max\set{c_i,0}
\end{equation}
and
\begin{equation}\label{jhjghg}
	w_i=-\frac{b_{i,i'}[x^{k_i}]}{\mu_i},\quad 
	\mu_i=\LC(a_{i,i}y_i\phi_s).
\end{equation}
where the bracket notation $f[x^k]$ refers to the coefficient of the term $x^k$ in $f$. Observe that $i'=(i+s)\bmod \gam$, and hence the map $i\mapsto i'$ is a permutation of $\set{0,1,\dots,\gam-1}$ and that  the integer $c_i$ is defined such that
\begin{equation}\label{asdfer}
	\gam c_i=\gd(d_{i',i'}\yb_{i'})-\rho(a_{i,i}y_i)-s.
\end{equation}

Now if $w_i=w$, let
\begin{equation}\label{laksnc}
	\tilde{g}_{i'}=\hat{g}_{i'},\quad
	\tilde{f}_i=\hat{f}_i
\end	{equation}
and if $w_i\neq w$ and $c_i>0$, let
\begin{equation}\label{cnksls}
	\tilde{g}_{i'}=\hat{f}_i,\quad
	\tilde{f}_i=x^{c_i}\hat{f}_i-\frac{\mu_i(w-w_i)}{\nu_{i'}^{(s)}}\hat{g}_{i'}	
\end{equation}
and if $w_i\neq w$ and $c_i\le 0$, let
\begin{equation}\label{cnskjf}
	\tilde{g}_{i'}=\hat{g}_{i'},\quad
	\tilde{f}_i=\hat{f}_i-\frac{\mu_i(w-w_i)}{\nu_{i'}^{(s)}}x^{-c_i}\hat{g}_{i'},
\end{equation}
where $\nu_i^{(s)}=\LC(d_{i,i})$. 

\begin{prop}\label{cmwdq}
The set $\tilde{B}=\set{\tilde{g}_i,\tilde{f}_i\mid 0\le i<\gam}$ is a Gr\"obner basis of $\tilde{I}$ with respect to $>_{s-1}$.
\end{prop}

\begin{IEEEproof}
Let $0\le i<\gam$.  We consider the pair
\[
	\begin{aligned}
	\hat{g}_{i'}&=\sum_{0\le j<\gam}c_{i',j}y_jz+\sum_{0\le j<\gam}d_{i',j}\yb_j+\sum_{0\le j<\gam}wc_{i',j}y_j\phi_s,\\
	\hat{f}_i&=\sum_{0\le j<\gam}a_{i,j}y_jz+\sum_{0\le j<\gam}b_{i,j}\yb_j+\sum_{0\le j<\gam}wa_{i,j}y_j\phi_s.	
	\end{aligned}
\]
By the assumption that $B^{(s)}$ is a Gr\"obner basis of $I_{v^{(s)}}$ with respect to $>_s$, we have for $0\le j<\gam$,
\[	
	\gd(d_{i',i'}\yb_{i'})>\gd_s(c_{i',j}y_jz)\ge\gd(wc_{i',j}y_j\phi_s)
\]
and for $0\le j<\gam$ with $j\neq i'$, $\gd(d_{i',i'}\yb_{i'})>\gd(d_{i',j}\yb_j)$. Therefore 
\[
	\LT_{s-1}(\hat{g}_{i'})=\LT(d_{i',i'}\yb_{i'}).
\]
Similarly we have for $0\le j<\gam$ with $j\neq i$,
\[
	\gd_s(a_{i,i}y_iz)>\gd_s(a_{i,j}y_jz)\ge\gd(wa_{i,j}y_j\phi_s)
\]
and for $0\le j<\gam$ with $j\neq i'$, $\gd_s(a_{i,i}y_iz)>\gd(b_{i,j}\yb_j)$ by the definition of $i'$ in \eqref{vnmskq}. Note that
\begin{equation}\label{ckmld}
	\gd_s(a_{i,i}y_iz)\ge\gd(b_{i,i'}\yb_{i'}+wa_{i,i}y_i\phi_s)
\end{equation}
where the inequality is strict if and only if $w=w_i$ by the definition of $w_i$ in \eqref{jhjghg}. Hence
if $w=w_i$, then $\LT_{s-1}(\hat{f}_i)=\LT_{s-1}(a_{i,i}y_iz)$ and if $w\neq w_i$, then $\LT_{s-1}(\hat{f}_i)=\LT(b_{i,i'}\yb_{i'}+wa_{i,i}y_i\phi_s)$.

Now we consider the set $\tilde{B}$ with respect to $>_{s-1}$. For the case that $w_i=w$, by \eqref{laksnc},
\begin{equation}\label{jfjwd}
	\begin{aligned}
	\LT_{s-1}(\tilde{g}_{i'})&=\LT_{s-1}(\hat{g}_{i'})=\LT(d_{i',i'}\yb_{i'}),\\
	\LT_{s-1}(\tilde{f}_i)&=\LT_{s-1}(\hat{f}_i)=\LT_{s-1}(a_{i,i}y_iz).
	\end{aligned}
\end{equation}
In the case that $w_i\neq w$ and $c_i>0$, we have \eqref{cnksls}. Observe that
\[
	\begin{aligned}
	\LT_{s-1}(x^{c_i}\hat{f}_i)&=x^{c_i}\LT(b_{i,i'}\yb_{i'}+wa_{i,i}y_i\phi_s),\\
	\LT_{s-1}(\hat{g}_{i'})&=\LT(d_{i',i'}\yb_{i'})
	\end{aligned}
\]
and by \eqref{ckmld} and \eqref{asdfer},
\[
	\gam c_i+\gd(b_{i,i'}\yb_{i'}+wa_{i,i}y_i\phi_s)
	=\gam c_i+\gd_s(a_{i,i}y_iz)
	=\gd(d_{i',i'}\yb_{i'}).
\]
Moreover
\[
	\begin{split}
	\LC_{s-1}(x^{c_i}\hat{f}_i)&=\LC(b_{i,i'}\yb_{i'}+wa_{i,i}y_i\phi_s)
	=-\mu_iw_i+\mu_iw\\
	&=\LC_{s-1}(\frac{\mu_i(w-w_i)}{\nu_{i'}^{(s)}}\hat{g}_{i'}).
	\end{split}
\]
This implies that there is a canceling of the leading coefficients in \eqref{cnksls}. Therefore, together with \eqref{ckmld}, we have
\begin{equation}\label{ccmmdd}
	\begin{aligned}
	\LT_{s-1}(\tilde{f}_i)&=\LT_{s-1}(x^{c_i}a_{i,i}y_iz),\\
	\LT_{s-1}(\tilde{g}_{i'})&=\LT_{s-1}(\hat{f}_i)=\LT(b_{i,i'}\yb_{i'}+wa_{i,i}y_i\phi_s).
	\end{aligned}
\end{equation}
For the case that $w_i\neq w$ and $c_i\le0$, we have \eqref{cnskjf}. By almost the same argument as above, we can show that
\begin{equation}\label{ckmslw}
	\LT_{s-1}(\tilde{g}_{i'})=\LT(d_{i',i'}\yb_{i'}),
	\quad
	\LT_{s-1}(\tilde{f}_i)=\LT_{s-1}(a_{i,i}y_iz).
\end{equation}
Finally it is clear that $\tilde{B}$ still generates the module $\tilde{I}$. From \eqref{jfjwd}, \eqref{ccmmdd}, and \eqref{ckmslw}, we see that $\tilde{B}$ is a Gr\"obner basis of $\tilde{I}$ with respect to $>_{s-1}$, by the criterion in Proposition~\ref{xnskw}.
\end{IEEEproof}

For the following, it is important to keep in mind that the values $w_i$, $c_i$ are determined only by $B^{(s)}$ and independent of $w$ although $\tilde{B}$ is clearly dependent on $w$.

\begin{lem}
Let $0\le i<\gam$. If $w_i\neq w$, then
\begin{equation}\label{zbjadd}
	\begin{split}
	\gd_{s-1}(\tilde{g}_{i'})&=\gd(d_{i',i'}\yb_{i'})-\gam \bar{c}_i,\\
	\gd_{s-1}(\tilde{f}_i)&=\gd_{s-1}(a_{i,i}y_iz)+\gam \bar{c}_i.
	\end{split}
\end{equation}
\end{lem}

\begin{IEEEproof}
Suppose $w_i\neq w$. Let us show the first equation. If $c_i>0$, then 
\[
	\begin{split}
	\gd_{s-1}(\tilde{g}_{i'})&=\gd_{s-1}(\hat{f}_i)
	=\gd(b_{i,i'}\yb_{i'}+wa_{i,i}y_i\phi_s)\\
	&=\gd_s(a_{i,i}y_iz)
	=\gd(d_{i',i'}\yb_{i'})-\gam c_i,
	\end{split}
\]	
by \eqref{ccmmdd}, \eqref{ckmld}, and \eqref{asdfer}. If $c_i\le 0$, then $\gd_{s-1}(\tilde{g}_{i'})=\gd(d_{i',i'}\yb_{i'})$ by \eqref{ckmslw}. The second equation is clear by \eqref{ccmmdd} and \eqref{ckmslw}.
\end{IEEEproof}

\begin{lem}\label{jfoqff}
For $i$ with $w_i\neq \gw_s$,
\[
	\rho(\epsilon_i)-\rho(a_{i,i}y_i)\ge\gam \bar{c}_i
\]
and
\[
	\min\set{\rho(\epsilon_i)+s,\gd(\eta_{i'})}\ge\gd(d_{i',i'}\yb_{i'}).
\]
\end{lem}

\begin{IEEEproof}
Suppose $w_i\neq\gw_s$. Then let us set $w=\gw_s$. Since $J_e(z-\gw_s\phi_s-\mu^{(s-1)})\subset I_{v^{(s)}}$, we have 
$J_e(z-\mu^{(s-1)})\subset\tilde{I}$. In particular, $\epsilon_i(z-\mu^{(s-1)})\in \tilde{I}$. Note that $\LT_{s-1}(\epsilon_i(z-\mu^{(s-1)}))=\LT_{s-1}(\epsilon_iz)$. As $\tilde{B}$ is a Gr\"obner basis of $\tilde{I}$ with respect to $>_{s-1}$ and $\ydeg(\epsilon_i)=i$, $\LT_{s-1}(\epsilon_iz)$ must be an $\F[x]$-multiple of $\LT_{s-1}(\tilde{f}_i)$. With \eqref{zbjadd}, this implies $\rho(\epsilon_i)\ge\rho(a_{i,i}y_i)+\gam \bar{c}_i$. Then by \eqref{asdfer}, 
\[
	\rho(\epsilon_i)-\rho(a_{i,i}y_i)\ge\gam \bar{c}_i\ge\gam  c_i=\gd(d_{i',i'}\yb_{i'})-\rho(a_{i,i}y_i)-s.
\]
Hence $\rho(\epsilon_i)+s\ge\gd(d_{i',i'}\yb_{i'})$. With Lemma \ref{xnndd}, this implies the second inequality. 
\end{IEEEproof}

\begin{lem}\label{kdlsef}
For $i$ with $w_i=\gw_s$,
\[
	\min\set{\rho(\epsilon_i)+s,\gd(\eta_{i'})}\ge\gd(d_{i',i'}\yb_{i'})-\gam \bar{c}_i
\]
\end{lem}

\begin{IEEEproof}
Suppose $w_i=\gw_s$. Then choose $w\in\F$ such that $w\neq \gw_s$. Since $J_e(z-\gw_s\phi_s-\mu^{(s-1)})\subset I_{v^{(s)}}$, we have 
\[
	J_e(z-(\gw_s-w)\phi_s-\mu^{(s-1)})\subset \tilde{I}.
\]
In particular, $\epsilon_i(z-(\gw_s-w)\phi_s-\mu^{(s-1)})\in \tilde{I}$. As $\gw_s-w\neq 0$, we have 
\[
	\LT_{s-1}(\epsilon_i(z-(\gw_s-w)\phi_s-\mu^{(s-1)}))=\LT((\gw_s-w)\epsilon_i\phi_s).
\]
By the definition of $i'$ and as $\tilde{B}$ is a Gr\"obner basis of $\tilde{I}$ with respect to $>_{s-1}$, $\LT((\gw_s-w)\epsilon_i\phi_s)$ must be an $\F[x]$-multiple of $\LT_{s-1}(\tilde{g}_{i'})$. Then $\rho(\epsilon_i)+s\ge\gd(d_{i',i'}\yb_{i'})-\gam \bar{c}_i$ by \eqref{zbjadd}. Finally, $\gd(\eta_{i'})\ge \gd(d_{i',i'}\yb_{i'})\ge \gd(d_{i',i'}\yb_{i'})-\gam \bar{c}_i$ by Lemma \ref{xnndd}.
\end{IEEEproof}

\begin{prop}
The condition
\[
	\sum_{0\le i<\gam}\max\set{\gd(\eta_{i'})-\rho(y_i)-s,\rho(\epsilon_i)-\rho(y_i)}>2\gam \wt(e)
\]
implies $\sum_{w_i=\gw_s}\bar{c}_i>\sum_{w_i\neq\gw_s}\bar{c}_i$.
\end{prop}

\begin{IEEEproof}
Lemmas \ref{jfoqff} and \ref{kdlsef} imply
\[
	\begin{split}
	\sum_{w_i=\gw_s}\gam \bar{c}_i
	&\ge\sum_{w_i=\gw_s}\gd(d_{i',i'}\yb_{i'})-\min\set{\rho(\epsilon_i)+s,\gd(\eta_{i'})}\\
	&\ge\sum_{0\le i<\gam}\gd(d_{i',i'}\yb_{i'})-\min\set{\rho(\epsilon_i)+s,\gd(\eta_{i'})}
	\end{split}
\]
and
\[
	\begin{split}
	\sum_{w_i\neq\gw_s}\gam \bar{c}_i
	&\le\sum_{w_i\neq\gw_s}\rho(\epsilon_i)-\rho(a_{i,i}y_i)\\
	&\le\sum_{0\le i<\gam}\rho(\epsilon_i)-\rho(a_{i,i}y_i).
	\end{split}
\]
Hence
\[
	\begin{split}
	&\sum_{w_i=\gw_s}\gam \bar{c}_i-\sum_{w_i\neq\gw_s}\gam \bar{c}_i\ge\sum_{0\le i<\gam}\rho(a_{i,i}y_i)+\gd(d_{i',i'}\yb_{i'})\\
	&\quad\quad-\min\set{2\rho(\epsilon_i)+s,\rho(\epsilon_i)+\gd(\eta_{i'})}\\
	&=\sum_{0\le i<\gam}\gd(\eta_{i'})+\rho(y_i)
	-\min\set{2\rho(\epsilon_i)+s,\rho(\epsilon_i)+\gd(\eta_{i'})}\\	
	&=\sum_{0\le i<\gam}\max\set{\gd(\eta_{i'})+\rho(y_i)-2\rho(\epsilon_i)-s,\rho(y_i)-\rho(\epsilon_i)}\\
	&=\sum_{0\le i<\gam}\max\set{\gd(\eta_{i'})-\rho(y_i)-s,\rho(\epsilon_i)-\rho(y_i)}-2\gam \wt(e)	
	\end{split}
\]
 where we used the equality
\[
	\begin{split}
 	&\sum_{0\le i<\gam}\rho(a_{i,i}y_i)+\gd(d_{i',i'}\yb_{i'})\\
	&=\sum_{0\le i<\gam}\gam\deg(a_{i,i})+\gam\deg(d_{i,i})+\rho(y_i)+\gd(\yb_i)\\
	&=\gam n+\sum_{0\le i<\gam}\rho(y_i)+\gd(\yb_i)
	=\sum_{0\le i<\gam}\gd(\eta_{i'})+\rho(y_i)
	\end{split}
\]
shown by Lemma \ref{lem_jjqqe} and Proposition \ref{camdzs}, and the equality
\[
	\sum_{0\le i<\gam}2(\rho(\epsilon_i)-\rho(y_i))=\sum_{0\le i<\gam}2\gam \xdeg(\epsilon_i)=2\gam \wt(e)
\]
shown by Proposition \ref{mdcad}.
\end{IEEEproof}

Let
\[
	\nu(s)=\frac{1}{\gam }\sum_{0\le i<\gam}\max\set{\gd(\eta_{i'})-\rho(y_i)-s,0}
\]
for $s\in\bL,s\le 0$. Then define 
\[
	\dLO=\min\set{\nu(s)\mid s\in\bL,s\le 0}.
\]

\begin{prop}\label{qwerrd}
The condition $\nu(s)>2\wt(e)$ implies
\[
	\sum_{w_i=\gw_s}\bar{c}_i>\sum_{w_i\neq\gw_s}\bar{c}_i.
\]
\end{prop}

\begin{IEEEproof}
Just note that $\rho(\epsilon_i)-\rho(y_i)\ge 0$ for $0\le i<\gam$.
\end{IEEEproof}

\begin{prop}
We have $\dLO\ge n-|G|$. 
\end{prop}

\begin{IEEEproof}
Note that
\[
	\begin{split}
	\nu(s)&=\frac{1}{\gam }\sum_{0\le i<\gam}\max\set{\gd(\eta_{i'})-\rho(y_i)-s,0}\\
	&\ge\frac{1}{\gam }\sum_{0\le i<\gam}(\gd(\eta_{i'})-\rho(y_i)-s)\\
	&=\frac{1}{\gam }\sum_{0\le i<\gam}(\gd(\eta_{i})-\rho(y_i))-s=n-|G|-s.
	\end{split}
\]
To show the last equality, pick any $f$ in $\Rbar$. Then
\[
	\begin{aligned}
	&\frac{1}{\gam }\sum_{0\le i<\gam}(\gd(\eta_i)-\rho(y_i))\\
	&=\frac{1}{\gam }\sum_{0\le i<\gam}(\gam\xdeg(\eta_i)+\gd(\yb_i)-\gd(y_if)+\gd(f))	\\
	&=\sum_{0\le i<\gam}\xdeg(\eta_i)-\sum_{0\le i<\gam}\xdeg(y_if)+\gd(f)\\
	&=\dim_\F \Rbar/J-\dim_\F \Rbar/(Rf)+\gd(f)\\
	&=n-|G|.
	\end{aligned}
\]
since
\[
	\begin{aligned}
	\dim_\F \Rbar/(Rf)&=\dim_\F\calL((s+\gd(f))Q+G)/\calL(sQ)f\\
	&=|G|+\gd(f)
	\end{aligned}
\]
for all large enough $s$.
\end{IEEEproof}

\subsection{Algorithm}\label{sec_qpwscks}

With the input $v\in\F^n$ the received vector, the algorithm below outputs the message $(\gw_{s_0},\gw_{s_1},\dots,\gw_{s_{k-1}})$ if $2\wt(e)<\dLO$.

\paragraph{Initialization} Let $N=\gd(h_v)$, and let $B^{(N)}$ be the Gr\"obner basis of $I_v$ with respect to $>_N$, 
\[
	\set{y_i(z-h_v),\eta_i\mid 0\le i<\gam}.
\]
Let $w_s=0$ for $s$ with $N<s\le 0,s\in\bL$. The following steps \textit{Pairing}, \textit{Voting}, and \textit{Rebasing} are iterated for $s$ decreasing from $N$ to $\lbar_0$.

\paragraph{Pairing} Suppose $B^{(s)}=\set{g_i^{(s)},f_i^{(s)}\mid 0\le i<\gam}$ is a Gr\"obner basis of $I_{v^{(s)}}$ with respect to $>_s$ where 
\[
	\begin{aligned}
	g_i^{(s)}&=\sum_{0\le j<\gam}c_{i,j}y_jz+\sum_{0\le j<\gam}d_{i,j}\yb_j\\
	f_i^{(s)}&=\sum_{0\le j<\gam}a_{i,j}y_jz+\sum_{0\le j<\gam}b_{i,j}\yb_j
	\end{aligned}
\]
and let $\nu_i^{(s)}=\LC(d_{i,i})$. For $0\le i<\gam$, let $i'=(i+s)\bmod\gam$, $k_i=\deg(a_{i,i})+(a_i+s-b_{i'})/\gam$,
and $c_i=\deg(d_{i',i'})-k_i$.

\paragraph{Voting}
If $s>0$ or $s\notin\bL$, then for $i$ with $k_i\ge 0$, let
\[
	w_i=-b_{i,i'}[x^{k_i}],\quad \mu_i=1
\]
and for $i$ with $k_i<0$, let $w_i=0, \mu_i=1$. Let $w=0$ in both cases.

If  $s\le 0$ and $s\in\bL$, then for each $i$, let
\[
	w_i=-\frac{b_{i,i'}[x^{k_i}]}{\mu_i},\quad \mu_i=\LC(a_{i,i}y_i\phi_s)
\]
and let $\bar{c}_i=\max\set{c_i,0}$, and let $w$ be the element of $\F$ with the largest
\[
	\sum_{w=w_i}\bar{c}_i,
\]
and let $w_s=w$.

\paragraph{Rebasing} For each $i$, do the following. If $w_i=w$, then let
\begin{equation}\label{fkfmvd}
	\begin{aligned}
	g_{i'}^{(s-1)}&=g_{i'}^{(s)}(z+w\phi_s)\\
	f_i^{(s-1)}&=f_i^{(s)}(z+w\phi_s)
	\end{aligned}
\end{equation}
and let $\nu_{i'}^{(s-1)}=\nu_{i'}^{(s)}$. 
If $w_i\neq w$ and $c_i>0$, then let
\begin{equation}\label{fkmksd}
	\begin{aligned}
	g_{i'}^{(s-1)}&=f_i^{(s)}(z+w\phi_s)\\
	f_i^{(s-1)}&=x^{c_i}f_i^{(s)}(z+w\phi_s)\\
	&\quad -\frac{\mu_i(w-w_i)}{\nu_{i'}^{(s)}}g_{i'}^{(s)}(z+w\phi_s)
	\end{aligned}	
\end{equation}
and let $\nu_{i'}^{(s-1)}=\mu_i(w-w_i)$.
If $w_i\neq w$ and $c_i\le 0$, then let
\begin{equation}\label{akfjcd}
	\begin{aligned}
	g_{i'}^{(s-1)}&=g_{i'}^{(s)}(z+w\phi_s)\\
	f_i^{(s-1)}&=f_i^{(s)}(z+w\phi_s)\\
	&\quad-\frac{\mu_i(w-w_i)}{\nu_{i'}^{(s)}}x^{-c_i}g_{i'}^{(s)}(z+w\phi_s)
	\end{aligned}
\end{equation}
and let $\nu_{i'}^{(s-1)}=\nu_{i'}^{(s)}$. Let $B^{(s-1)}=\set{g_i^{(s-1)},f_i^{(s-1)}\mid 0\le i<\gam}$.

\paragraph{Output}
After the iterations, output the recovered message $(w_{s_0},w_{s_1},\dots,w_{s_{k-1}})$.

We now give an overview of the algorithm. Note that the decoding algorithm is in one of two phases while $s$ decreases from $N$ to $\lbar_0$. The first phase is when $s>0$ or $s\notin\bL$, and the second phase is when $s\le 0, s\in\bL$. In the first phase,  the Gr\"obner basis $B^{(s)}$ of $I_{v^{(s)}}$ with respect to $>_s$ is updated such that $B^{(s-1)}$ is a Gr\"obner basis of $I_{v^{(s-1)}}$ with respect to $>_{s-1}$ where
\[
	v^{(s-1)}=v^{(s)}.
\]
In the second phase, the algorithm determines $w_s$ by majority voting and updates $B^{(s)}$ such that $B^{(s-1)}$ is a Gr\"obner basis of $I_{v^{(s-1)}}$ with respect to $>_{s-1}$ where  
\[
	v^{(s-1)}=v^{(s)}-\ev(w_s\phi_s).
\]
When the algorithm terminates, $w_s$ are determined for all $s\in \bL, s \le 0$.

\begin{prop}\label{cmskqe}
For $N\ge s\ge\lbar_0$, the set $B^{(s)}$ is a Gr\"obner basis of $I_{v^{(s)}}$ with respect to $>_s$.
\end{prop}

\begin{IEEEproof}
This is proved by induction on $s$. For $s=N$, this is true by \eqref{cjsjsw}. Now our induction assumption is that this is true for $s$. In the second phase, we already saw in Proposition \ref{cmwdq} that $B^{(s-1)}$ is a Gr\"obner basis of $I_{v^{(s-1)}}$. So it remains to consider the first phase. The proof for this case is similar to that of Proposition \ref{cmwdq}.

Suppose $s>0$ or $s\notin\bL$. Let $0\le i<\gam$. Recall
\[
	\begin{aligned}
	g_{i'}^{(s)}&=\sum_{0\le j<\gam}c_{i',j}y_jz+\sum_{0\le j<\gam}d_{i',j}\yb_j\\
	f_i^{(s)}&=\sum_{0\le j<\gam}a_{i,j}y_jz+\sum_{0\le j<\gam}b_{i,j}\yb_j
	\end{aligned}
\]
By the induction assumption, we have for $0\le j<\gam$,
\[	
	\gd(d_{i',i'}\yb_{i'})>\gd_s(c_{i',j}y_jz)=\rho(c_{i',j}y_j)+s
\]
and for $0\le j<\gam$ with $j\neq i'$, $\gd(d_{i',i'}\yb_{i'})>\gd(d_{i',j}\yb_j)$. Therefore $\LT_{s-1}(g_{i'}^{(s)})=\LT(d_{i',i'}\yb_{i'})$.
Similarly, by the induction assumption, we have for $0\le j<\gam$ with $j\neq i$, $\gd_s(a_{i,i}y_iz)>\gd_s(a_{i,j}y_jz)$ and for $0\le j<\gam$ with $j\neq i'$, $\gd_s(a_{i,i}y_iz)>\gd(b_{i,j}\yb_j)$.

Note that
\begin{equation}\label{njlkih}
	\gd_s(a_{i,i}y_iz)\ge\gd(b_{i,i'}\yb_{i'})
\end{equation}
where the inequality is strict except when $\rho(a_{i,i}y_i)+s\in\bL$ and $b_{i,i'}[x^{k_i}]\neq 0$. 
Recall that $w_i=0$ if and only if $\rho(a_{i,i}y_i)+s\notin\bL$ or $\rho(a_{i,i}y_i)+s\in\bL$ but $b_{i,i'}[x^{k_i}]=0$. Therefore  if $w_i=0$, then $\LT_{s-1}(f_i^{(s)})=\LT_{s-1}(a_{i,i}y_iz)$ and if $w_i\neq 0$, then $\LT_{s-1}(f_i^{(s)})=\LT(b_{i,i'}\yb_{i'})$.

Now in the case when $w_i=0$, by \eqref{fkfmvd} and \eqref{njlkih},
\[
	\begin{split}
	\LT_{s-1}(g_{i'}^{(s-1)})&=\LT_{s-1}(g_{i'}^{(s)})=\LT(d_{i',i'}\yb_{i'}),\\
	\LT_{s-1}(f_i^{(s-1)})&=\LT_{s-1}(f_i^{(s)})=\LT_{s-1}(a_{i,i}y_iz).
	\end{split}
\]
In the case when $w_i\neq 0$ and $c_i>0$, by \eqref{fkmksd},
\[
	g_{i'}^{(s-1)}=f_i^{(s)},
	\quad
	f_i^{(s-1)}=x^{c_i}f_i^{(s)}+\frac{\mu_iw_i}{\nu_{i'}^{(s)}}g_{i'}^{(s)}.
\]
Observe that
\[
	\begin{gathered}
	\LT_{s-1}(x^{c_i}f_i^{(s)})=x^{c_i}\LT(b_{i,i'}\yb_{i'}),
	\quad
	\LT_{s-1}(g_{i'}^{(s)})=\LT(d_{i',i'}\yb_{i'}),\\
	\gam c_i+\gd(b_{i,i'}\yb_{i'})
	=\gam c_i+\gd_s(a_{i,i}y_iz)
	=\gd(d_{i',i'}\yb_{i'}),
	\end{gathered}
\]
and by the equality in \eqref{njlkih},
\[
	\LC_{s-1}(x^{c_i}f_i^{(s)})=\LC(b_{i,i'}\yb_{i'})=-\mu_iw_i
	=-\LC_{s-1}(\frac{\mu_iw_i}{\nu_{i'}^{(s)}}g_{i'}^{(s)}).
\]
This implies $\LT_{s-1}(f_i^{(s-1)})=\LT_{s-1}(x^{c_i}a_{i,i}y_iz)$.

Finally in the case when $w_i\neq 0$ and $c_i\le0$, by \eqref{akfjcd},
\[
	g_{i'}^{(s-1)}=g_{i'}^{(s)},
	\quad
	f_i^{(s-1)}=f_i^{(s)}+\frac{\mu_iw_i}{\nu_{i'}^{(s)}}x^{-c_i}g_{i'}^{(s)}.
\]
Then we can show that $\LT_{s-1}(f_i^{(s-1)})=\LT_{s-1}(a_{i,i}y_iz)$ by the same argument as when $c_i>0$.

Hence all in all the set $B^{(s-1)}$ is a Gr\"obner basis of $I_{v^{(s-1)}}$ with respect to $>_{s-1}$ also in the first phase.
\end{IEEEproof}

\begin{prop}
If $2\wt(e)<\dLO$, then $w_s=\gw_s$ for all $s\in\bL,s\le 0$. Hence
\[
	\sum_{s\in\bL,s\le 0} w_s\phi_s=\mu.
\]
\end{prop}

\begin{IEEEproof}
If $2\wt(e)<\dLO$, then Propositions \ref{qwerrd} and \ref{cmskqe} imply $w_s=\gw_s$ for all $s\in\bL,s\le 0$. 
\end{IEEEproof}

\subsection{Complexity}\label{sec_jdmxzd}

Recall that the main data with which the decoding algorithm works is essentially $2\gam\times 2\gam$ array of polynomials in $\F[x]$ that represents $B^{(s)}$. Each of the $2\gam$ rows of the array are again viewed as pairs of vectors in $\F[x]^\gam$. To optimize the speed complexity of the algorithm, it is necessary to precompute and store required information as vectors in $\F[x]^{\gam}$ before the error correction processing for the received vector $v$ begins. 

For the \textit{Initialization} step, we precompute $h_i$ for $1\le i\le n$ and $\eta_i$ for $0\le i<\gam$ in the vector form. Then for given $v$, $h_v$ is computed just as an $\F$-linear combination of the vectors. Thus the setup of the initial Gr\"obner basis $B^{(N)}$ is straightforward.

In the \textit{Rebasing} step, the most intensive computation is the substitution of $z$ with $z+w\phi_s$. As $\phi_s$ is in the form $x^k\yb_i$, the computation is facilitated if $y_i\yb_j$ for $0\le i,j<\gam$ is precomputed in the vector form. The necessity of the precomputation of $y_i\yb_j$ was first noted in \cite{geil2012c} for the case of general one-point codes.

If the output of the algorithm at the \textit{Output} step should be the corrected codeword, say, under systematic encoding, then precomputation of the vectors $\ev(\phi_{s_i})$ in $\F^n$ for $0\le i\le k-1$, essentially the generator matrix of the code $C$, would be necessary.

\begin{prop}
Lagrange basis polynomial $h_i$ can be chosen such that the maximum degree of the polynomials in the vector form of $h_i$ is bounded by
\[
	N_{h}=\fl{(n+2g-1)/\gam}.
\]
\end{prop}

\begin{IEEEproof}
By the Riemann-Roch, we can choose $h_i$ in
\[
	\calL(sQ+G+P_i-\sum_{1\le j\le n}P_j)/\calL(sQ+G-\sum_{1\le j\le n}P_j)
\]
if $s+|G|-n=2g-1$, and hence $\gd(h_i)\le n-|G|+2g-1$. Suppose $h_i=\sum_{0\le j<\gam} h_{ij}\yb_j$ with $h_{ij}\in\F[x]$. Then 
\[
	\gam\deg(h_{ij})+\gd(\yb_j)\le n-|G|+2g-1
\]
Since $\gd(\yb_j)\ge -|G|$, we have $	\deg(h_{ij})\le (n+2g-1)/\gam$.
\end{IEEEproof}

\begin{prop}
The maximum degree of the polynomials in the vector form of $\eta_i$ is bounded by
\[
	N_{\eta}=\fl{(n+g)/\gam}.
\]
\end{prop}

\begin{IEEEproof}
Since $\dim_\F \Rbar/J=n$, there can be no more than $n$ monomials preceding $\LM(\eta_i)$, which implies $\gd(\eta_i)\le\lbar_n$. Recall that $\gL+\lbar_0\subset\bL$. Therefore $\lbar_n\le \lbar_0+n+g$. Suppose that $\eta_i=\sum_{0\le j<\gam}\eta_{ij}\yb_j$ with $\eta_{ij}\in\F[x]$. Then 
\[
	\gam\deg(\eta_{ij})+\gd(\yb_j)\le\gd(\eta_i)\le \lbar_0+n+g.
\]
Since $\gd(\yb_j)\ge \lbar_0$, we have $\deg(\eta_{ij})\le (n+g)/\gam$.
\end{IEEEproof}

\begin{prop}
The maximum degree of the polynomials in the  $2\gam\times2\gam$ array during an execution is bounded by 
\[
	N_{\mathrm{deg}}=1+\fl{(n+4g-2)/\gam}
\]
if $g>0$. If $g=0$, then it is bounded by $n$.
\end{prop}

\begin{IEEEproof}
First observe that the behavior of the algorithm is such that the maximum of $\gd(f)$ for $f\in B^{(s)}$ is monotonically decreasing through the iterations. So it suffices to consider $\gd(\eta_i)$ and $\gd(y_ih_v)$ in the initial basis $B^{(N)}$. Since $\gd(h_i)\le n-|G|+2g-1$ and $\rho(y_i)=a_i\le 2g+\gam-1$ by the definition of $a_i$, we have
\[
	\gd(y_ih_v)\le \gam+n-|G|+4g-2
\]
On the other hand, $\gd(\eta_i)\le\lbar_0+n+g$. Hence during the execution, we have for $f\in B^{(s)}$,
\[
	\gd(f)=\max\set{\gam+n-|G|+4g-2,\lbar_0+n+g},
\]
from which we deduce that the maximum degree of the polynomials in the array is bounded by
\[
	\max\set{1+(n+4g-2)/\gam,(n+g)/\gam},
\]
where the former is larger if $g>0$. If $g=0$, the latter is larger, and is $n$.
\end{IEEEproof}

\begin{prop}
The number of iterations is at most
\[
	N_{\mathrm{iter}}=n+2g,
\]
\end{prop}

\begin{IEEEproof}
The algorithm iterates from $\gd(h_v)$ to $\lbar_0$. Since $\gd(h_v)\le n-|G|+2g-1$ and $\lbar_0\ge -|G|$, the number of iterations is at most $\gd(h_v)-\lbar_0+1\le n+2g$.
\end{IEEEproof}

\begin{prop}\label{kcmsff}
If $g>0$, an execution of the decoding algorithm takes $O((n+4g)(n+2g)g)$ multiplications. For $g=0$, it takes $O(n^2)$ multiplications. The implicit constant is absolute.
\end{prop}

\begin{IEEEproof}
For the first phase iteration, the update for each pair of the upper and lower rows of the array takes $O(n+4g+\gam)$ multiplications. Hence for the whole array, it takes $O((n+4g+\gam)\gam)$. 
For the second phase iteration, note that the maximum degree of the polynomials in the vector form of $y_i\yb_j$ is $(4g+2\gam-2)/\gam$. Hence the substitution operation for each row takes $O((n+4g)(2g+\gam)/\gam)$. For the whole array, it is $O((n+4g)(2g+\gam))$.

If $g>0$, then $\gam\le g$, so an iteration in either of first phase and second phase takes $O((n+4g)g)$ multiplications. Thus for $N_{\mathrm{iter}}$ number of iterations, it takes $O((n+4g)(n+2g)g)$ multiplications. On the other hand, $\gam=1$ for $g=0$. Finally the dominant part of the computation of the initial basis $B^{(N)}$ is the computation of $h_v$, which takes $O(n(n+2g))$ multiplications.  
\end{IEEEproof}

\section{Examples}\label{iqwkdq}

In this section, we give some explicit examples illustrating our decoding algorithm. We implemented the algorithm in Magma \cite{magma1997}. In particular, for the computation of $y_i$ and $\yb_i$, He\ss' algorithm \cite{hess2002} is heavily used as implemented in Magma. For the computation of $\eta_i$, we used a custom FGLM algorithm \cite{faugere1993}.

\subsection{Two-Point Hermitian Code}

Let $X$ be the Hermitian curve defined by
\[
	\y^3+\y=\x^4
\]
over $\F_9=\F_{3}(\ga)$ with $\ga^2-\ga-1=0$. The genus of $X$ is $3$. Let $G=-O+18Q$ where $O$ is the origin and $Q$ is the unique point at infinity. Except $O$ and $Q$, there are $26$ rational points
\[
	\begin{gathered}
	 ( 0, \ga^2 ), ( 0, \ga^6 ), ( 1, 2 ), ( 1, \ga ), ( 1, \ga^3 ), ( 2, 2 ), ( 2, \ga ),\\
	 ( 2, \ga^3 ),( \ga, 1 ), ( \ga, \ga^7 ), ( \ga, \ga^5 ), ( \ga^2, 2 ), ( \ga^2, \ga ),(\ga^2, \ga^3 ),\\
	 ( \ga^7, 1 ),( \ga^7, \ga^7 ), ( \ga^7, \ga^5 ), ( \ga^5, 1 ), ( \ga^5, \ga^7 ), ( \ga^5, \ga^5 ), \\
	 ( \ga^3, 1 ),( \ga^3, \ga^7 ), ( \ga^3, \ga^5 ), ( \ga^6, 2 ), ( \ga^6, \ga ), ( \ga^6, \ga^3 ).
	\end{gathered}
\]
Then the AG code $C=C_\calL(D,G)$ is a $[26,15,9]$ linear code over $\F_9$.

The Weierstrass semigroup at $Q$ is 
\[
	\gL=\set{0,3,4,6,7,8,9,\dots}.
\]
So $\gam=3$, and we take $x=\x$. The $\F[x]$-basis of $R$ is
\[
	\begin{aligned}
	y_0&=1, &\rho(y_0)=0,\\
	y_1&=\y, &\rho(y_1)=4, \\
	y_2&=\y^2, &\rho(y_2)=8.
	\end{aligned}
\]
On the other hand,
\[	
	\begin{split}
	\bL&=\{-15,-14,-12,-11,-10,-9,-8,-7,-6,-5,-4,\\
	&\quad-3,-2,-1,0,1,2,3,\dots\}
	\end{split},
\]
and the $\F[x]$-basis of $\Rbar$ is
\[
	\begin{aligned}
	\yb_0&=\x, &\gd(\yb_0)=-15,\\
	\yb_1&=\y,&\gd(\yb_1)=-14,\\
	\yb_2&=\y^2,&\gd(\yb_2)=-10.
	\end{aligned}
\]
The $\F[x]$-basis of $J$ is
\[
	\begin{aligned}
	\eta_0&=(x^8-1)\yb_0,\\
	\eta_1&=(x^9-x)\yb_1,\\
	\eta_2&=(x^9-x)\yb_2.
	\end{aligned}
\]
Using the above data, we can compute $\dLO=9$ since
\[
	\begin{array}{cccc}
	s &  \nu(s) \\
	\hline
	0&9\\
	-1&10\\
	-2&11\\
	-3&12\\
	-4&13\\
	-5&14\\
	-6&15\\
	-7&16\\
	\end{array}
	\quad
	\begin{array}{ccc}
	s &  \nu(s) \\
	\hline
	-8&17\\
	-9&18\\
	-10&19\\
	-11&20\\
	-12&21\\
	-14&23\\
	-15&24\\
	\\
	\end{array}
\]
The Lagrange basis for $\Rbar$ is
\[
	\begin{aligned}
	h_1&=(-\x^8 + 1)\y^2 + (\ga^6\x^8 + \ga^2)\y,\\
	h_2&=(-\x^8 + 1)\y^2 + (\ga^2\x^8 + \ga^6)\y,\\
	\vdots\\
	h_{26}&=(-\x^8 + \ga^2\x^7 + \cdots + \ga^6\x)\y^2 \\
	&\quad+ (\ga^7\x^8 + \ga^5\x^7 + \cdots + \ga\x)\y \\
	&\quad+ \ga\x^8 + \ga^7\x^7 + \cdots + \ga^3\x.
    	\end{aligned}
\]

Now suppose that the received vector is
\[
	\begin{split}
	v&=(0,0,0,0,\ga^2,-1,0,0,0,0,0,0,0,0,0,0,0,0,\\
	&\quad \ga^3,0,0,0,0,0,-1,0)\in\F_9^{26}.
	\end{split}
\]
Then the six generators of the module $I_v$ are
\[
	\begin{aligned}
	g_0&=\eta_0,\\
	g_1&=\eta_1,\\
	g_2&=\eta_2,\\
	f_0&=y_0(z-h_v),\\
	f_1&=y_1(z-h_v),\\
	f_2&=y_2(z-h_v),
	\end{aligned}
\]
where
\[
	\begin{split}
	h_v&=(\ga^7\x^7 + 2\x^6 + \ga^3\x^5 + \ga^7\x^4 + \ga^2\x^3 + \ga^7\x^2 + \ga\x)\y^2 \\
	&\quad+ (\ga^2\x^8 + \ga^2\x^7 + \ga^6\x^5 + 2\x^3 + \x^2 + \ga^7\x)\y \\
	&\quad+ \x^8 + \ga^6\x^7 + \ga^5\x^5 + 2\x^3 + \ga^6\x^2 + \ga^2\x.
	\end{split}
\]

Since $N=\gd(h_v)=11$, the initial basis of $I_v$ in \eqref{znxkdx} is a Gr\"obner basis with respect to $>_{11}$. Then we move on to the main iterative steps. In the first \textit{Pairing} and \textit{Voting} steps, the following data is computed:
\[
\begin{array}{ccrr}
\multicolumn{4}{c}{s=11}\\
\hline
 i&{i'}&c_i&w_i \\
 \hline
0&2&2&\ga^7\\
1&0&-2&\ga^7\\
2&1&-2&\ga^7
\end{array}
\]
In the \textit{Rebasing} step, the basis is updated to \eqref{jjskdf}, which is a Gr\"obner basis with respect to $>_{10}$. Similar updates are iterated until $s$ reaches to $0$. The Gr\"obner basis of $I_{v^{(0)}}$ with respect to $>_0$ is \eqref{jsfggf}. Now that $s\in\bL,s\le 0$, the algorithm goes into the second phase in which majority voting takes place. We listed in \eqref{xbajde} the data computed in the \textit{Pairing} and \textit{Voting} steps. For example, for $s=0$,  the winner $w$ in the voting is $0$. The basis after the final iteration is \eqref{cmskq}. Note that the recovered message is $0\in\F^{14}$.

\begin{figure*}[ht]
\begin{gather}
\label{znxkdx}
\begin{array}{r*{6}{|r}}
 &\makebox[50pt][r]{$y_2z$}&\makebox[50pt][r]{$y_1z$}&\makebox[50pt][r]{$y_0z$}
 &\makebox[50pt][r]{$\yb_2$}&\makebox[50pt][r]{$\yb_1$}&\makebox[50pt][r]{$\yb_0$}\\
\hline
g_0& &  &  &  &  & x^8-1 \\
g_1& &  &  &  &x^9-x& \\
g_2& &  &  &x^9-x&  & \\
f_0& &  & 1 &\ga^3x^7+\cdots&\ga^6x^8+\cdots & -x^7+\cdots \\
f_1& & 1&  & \ga^6x^8+\cdots &-x^8+\cdots&\ga^3x^{10}+\cdots\\
f_2& 1&  &  &-x^8+\cdots &\ga^3x^{11}+\cdots& \ga^6x^{11}+\cdots
\end{array}
\\[2ex]
\label{jjskdf}
\begin{array}{r*{6}{|r}}
 &\makebox[50pt][r]{$y_2z$}&\makebox[50pt][r]{$y_1z$}&\makebox[50pt][r]{$y_0z$}
 &\makebox[50pt][r]{$\yb_2$}&\makebox[50pt][r]{$\yb_1$}&\makebox[50pt][r]{$\yb_0$}\\
\hline
g_0& &  &  &  &  & x^8-1 \\
g_1& &  &  &  &x^9-x& \\
g_2& &  & 1 &\ga^3x^7+\cdots&\ga^6x^8+\cdots & -x^7+\cdots\\
f_0& &  & x^2 &x^8+\cdots&\ga^6x^{10}+\cdots & -x^9+\cdots \\
f_1& & 1&  & \ga^6x^8+\cdots &-x^8+\cdots&x^9+\cdots\\
f_2& 1&  &  &-x^8+\cdots &x^{10}+\cdots& \ga^6x^{11}+\cdots
\end{array}
\\[2ex]
\label{jsfggf}
\begin{array}{r*{6}{|r}}
 &\makebox[50pt][r]{$y_2z$}&\makebox[50pt][r]{$y_1z$}&\makebox[50pt][r]{$y_0z$}
 &\makebox[50pt][r]{$\yb_2$}&\makebox[50pt][r]{$\yb_1$}&\makebox[50pt][r]{$\yb_0$}\\
\hline
g_0& &  &  &  &  & x^8-1 \\
g_1& &1  &\ga^7x+\ga & -x^6+\cdots &x^8+\cdots& \ga^3x^7+\cdots\\
g_2& &  & 1 &\ga^3x^7+\cdots&\ga^6x^8+\cdots & -x^7+\cdots\\
f_0& & 1 & x^2+\cdots &  &  &  \\
f_1& & x+\ga& \ga^7x^2+\cdots & \ga x^5+\cdots &\ga^5x^7+\cdots&\ga^6x^7+\cdots\\
f_2& 1& \ga^7 & \ga^7x+\ga^7  & \ga^2x^5+\cdots &\ga^6x^7+\cdots& \ga^7x^7+\cdots
\end{array}
\\[2ex]
\label{cmskq}
\begin{array}{r*{6}{|r}}
 &\makebox[50pt][r]{$y_2z$}&\makebox[50pt][r]{$y_1z$}&\makebox[50pt][r]{$y_0z$}
 &\makebox[50pt][r]{$\yb_2$}&\makebox[50pt][r]{$\yb_1$}&\makebox[50pt][r]{$\yb_0$}\\
\hline
g_0& &  &  &  &  & x^8-1 \\
g_1& &x+\ga &\ga^7x^2+\cdots &\ga x^5+\cdots &\ga^5x^7+\cdots& \ga^6x^7+\cdots\\
g_2& &  & 1 &\ga^3x^7+\cdots&\ga^6x^8+\cdots & -x^7+\cdots\\
f_0& & 1 & x^2+\cdots &  &  &  \\
f_1& & x^2+\cdots& \ga^7x^3+\cdots & &&\\
f_2& 1& \ga^5x+1 & -x^2+\cdots  &&& 
\end{array}
\end{gather}
\hrulefill
\vspace*{4pt}
\end{figure*}

\begin{figure*}[ht]
\begin{equation}\label{xbajde}
\begin{gathered}
\begin{array}{ccrr}
\multicolumn{4}{c}{s=0}\\
\hline
i&{i'}&c_i&w_i \\
\hline
0&0&1&0\\
1&1&1&\ga\\
2&2&1&0
\end{array}
\quad
\begin{array}{ccrr}
\multicolumn{4}{c}{s=-1}\\
\hline
i&{i'}&c_i&w_i \\
\hline
0&2&2&0\\
1&0&0&\ga^2 \\
2&1&0&\ga^2
\end{array}
\quad
\begin{array}{ccrr}
\multicolumn{4}{c}{s=-2}\\
\hline
i&{i'}&c_i&w_i \\
\hline
0&1&1&0\\
1&2&1&0\\
2&0&1&0
\end{array}
\quad
\begin{array}{ccrr}
\multicolumn{4}{c}{s=-3}\\
\hline
i&{i'}&c_i&w_i \\
\hline
0&0&2&0\\
1&1&0&\ga^5\\
2&2&2&0
\end{array}
\quad
\begin{array}{ccrr}
\multicolumn{4}{c}{s=-4}\\
\hline
 i&{i'}&c_i&w_i \\
 \hline
0&2&3&0\\
1&0&1&0\\
2&1&1&0
\end{array}
\\
\begin{array}{ccrr}
\multicolumn{4}{c}{s=-5}\\
\hline
i&{i'}&c_i&w_i \\
\hline
0&1&2&0\\
1&2&2&0\\
2&0&2&0
\end{array}
\quad
\begin{array}{ccrr}
\multicolumn{4}{c}{s=-6}\\
\hline
i&{i'}&c_i&w_i \\
\hline
0&0&3&0\\
1&1&1&0\\
2&2&3&0
\end{array}
\quad
\begin{array}{ccrr}
\multicolumn{4}{c}{s=-7}\\
\hline
i&{i'}&c_i&w_i \\
\hline
0&2&4&0\\
1&0&2&0\\
2&1&2&0
\end{array}
\quad
\begin{array}{ccrr}
\multicolumn{4}{c}{s=-8}\\
\hline
i&{i'}&c_i&w_i \\
\hline
0&1&3&0\\
1&2&3&0\\
2&0&3&0
\end{array}
\quad
\begin{array}{ccrr}
\multicolumn{4}{c}{s=-9}\\
\hline
i&{i'}&c_i&w_i \\
\hline
0&0&4&0\\
1&1&2&0\\
2&2&4&0
\end{array}
\\
\begin{array}{ccrr}
\multicolumn{4}{c}{s=-10}\\
\hline
i&{i'}&c_i&w_i \\
\hline
0&2&5&0\\
1&0&3&0\\
2&1&3&0
\end{array}
\quad
\begin{array}{ccrr}
\multicolumn{4}{c}{s=-11}\\
\hline
i&{i'}&c_i&w_i \\
\hline
0&1&4&0\\
1&2&4&0\\
2&0&4&0
\end{array}
\quad
\begin{array}{ccrr}
\multicolumn{4}{c}{s=-12}\\
\hline
i&{i'}&c_i&w_i \\
\hline
0&0&5&0\\
1&1&3&0\\
2&2&5&0
\end{array}
\quad
\begin{array}{ccrr}
\multicolumn{4}{c}{s=-14}\\
\hline
i&{i'}&c_i&w_i \\
\hline
0&1&5&0\\
1&2&5&0\\
2&0&5&0
\end{array}
\quad
\begin{array}{ccrr}
\multicolumn{4}{c}{s=-15}\\
\hline
i&{i'}&c_i&w_i \\
\hline
0&0&6&0\\
1&1&4&0\\
2&2&6&0
\end{array}
\end{gathered}
\end{equation}
\hrulefill
\vspace*{4pt}
\end{figure*}

\subsection{Two-Point Code on the Klein Quartic}

The Klein quartic over $\F_8$ is defined by the equation
\[
	\y^3+\x^3\y+\x=0.
\]
The genus of the curve is $3$. The curve has $24$ rational points including two points $Q_1=[0\!:\!1\!:\!0]$, $Q_2=[1\!:\!0\!:\!0]$ at infinity. Let $G=-Q_1+19Q_2$ and $Q=Q_1$.
The Weierstrass semigroup at $Q$ is 
\[
	\gL=\set{0,3,5,6,7,8,\dots}.
\]
Hence $\gam=3$, and we take $x=\y$. Then
\[
	\begin{aligned}
	y_0&=1, &\rho(y_0)=0,\\
	y_1&=\y\x^2,&\rho(y_1)=7,\\
	y_2&=\y\x,&\rho(y_2)=5.
	\end{aligned}
\]
We find that
\[	
	\bL=\{-17,-14,-13,-12,-11,\dots\}
\]
and
\[
	\begin{aligned}
	\yb_0&=\x^2/\y^8+\x/\y^5, &\gd(\yb_0)=-12,\\
	\yb_1&=\x/\y^9+1/\y^6,&\gd(\yb_1)=-17,\\
	\yb_2&=\x^2/\y^6,&\gd(\yb_2)=-13.
	\end{aligned}
\]
The $\F[x]$-basis of $J$ is
\[
	\begin{aligned}
	\eta_0&=(x^7+1)\yb_0,\\
	\eta_1&=(x^7+1)\yb_1,\\
	\eta_2&=(x^8+x)\yb_2.
	\end{aligned}
\]
Note that we have $\dLO=5 $ since
\[
	\begin{array}{cc}
	s & \nu(s) \\
	\hline
	0&5\\
	-1&5\\
	-2&6\\
	-3&7\\
	-4&8\\
	-5&9\\
	-6&10\\
	-7&11\\
	\end{array}
	\quad
	\begin{array}{cc}
	s &  \nu(s) \\
	\hline
	-8&12\\
	-9&13\\
	-10&14\\
	-11&15\\
	-12&16\\
	-13&17\\
	-14&18\\
	-17&21\\
	\end{array}	
\]
Indeed the code $C=C_\calL(D,G)$ is $[22,16,5]$ linear code over $\F_9$. So the decoding algorithm corrects errors up to half of the minimum distance.

Now let us see what happens if we take $Q=Q_2$. As the code $C_\calL(D,G)$ itself is independent of the choice of $Q$, we obtain the same linear code. Incidentally $\gL$ does not change, and we have the same $\gam=3$, but we should take $x=\x/\y$, and
\[
	\begin{aligned}
	y_0&=1, &\rho(y_0)=0,\\
	y_1&=\x/\y^3,&\rho(y_1)=7,\\
	y_2&=\x/\y^2,&\rho(y_2)=5.
	\end{aligned}
\]
On the other hand, we have different
\[	
	\bL=\{-16,-14,-13,-12,-11,\dots\}
\]
and
\[
	\begin{aligned}
	\yb_0&=\x/\y^3,&\gd(\yb_0)=-12,\\
	\yb_1&=\x/\y^2,&\gd(\yb_1)=-14,\\
	\yb_2&=\x/\y,&\gd(\yb_2)=-16.
	\end{aligned}
\]
This time the $\F[x]$-basis $J$ is
\[
	\begin{aligned}
	\eta_0&=(x^8+x)\yb_0,\\
	\eta_1&=(x^7+1)\yb_1,\\
	\eta_2&=(x^7+1)\yb_2,
	\end{aligned}
\]
and 
\[
	\begin{array}{cc}
	s & \nu(s) \\
	\hline
	0&4\\
	-1&5\\
	-2&6\\
	-3&7\\
	-4&8\\
	-5&9\\
	-6&10\\
	-7&11\\
	\end{array}
	\quad
	\begin{array}{cc}
	s & \nu(s) \\
	\hline
	-8&12\\
	-9&13\\
	-10&14\\
	-11&15\\
	-12&16\\
	-13&17\\
	-14&18\\
	-16&20\\
	\end{array}	
\]
Thus we have $\dLO=4$ this time. This example shows that the performance of our decoding algorithm indeed depends on the choice of $Q$ in a subtle way.

\subsection{Two-Point Code on a Suzuki Curve}

Let us consider the Suzuki curve
\[
	\y^8-\y=\x^2(\x^8-\x)
\]
over $\F_8$. The genus of the curve is $g=14$. This curve has $65$ rational points including one cusp at infinity. Let $G=15O+24Q$ where $O$ is the origin and $Q$ is the unique place at the cusp. Let $D$ be the sum of other $63$ rational points. Then the code $C_\calL(D,G)$ is a $[63,26,\ge 25]$ linear code over $\F_8$ with the best known minimum distance for codes of length $63$ and dimension $26$ over $\F_8$ \cite{matthews2004}. We have $\dLO=25$. 

Recall that $n=63$, $g=14$, and $\gam=8$. The maximum degree of the polynomials in the vector forms of $h_i$ is $7$ ($N_h=11$). The maximum degree of the polynomials in the vector forms of $\eta_i$ is $8$ ($N_\eta=9$). In an experiment with $10^5$ instances of decoding random errors of weight $12$, the decoder performed at most $82$ ($N_{\mathrm{iter}}=91$) iterations with an $16\times 16$ matrix of univariate polynomials at most $13$ ($N_{\mathrm{deg}}=16$) degree over $\F_8$. It took 0.0397 second to decode one instance on Macbook Pro, taking $O(151606)$ multiplications according to Proposition \ref{kcmsff}.

\subsection{Two-Point Reed-Solomon Code}

The projective line over $\F_{64}$ is a curve with genus $0$ whose function field is the rational function field $\F_{64}(x)$. It has $65$ rational points including the point at infinity. Let $G=-O+39Q$ where $O$ is the origin and $Q$ is the point at infinity. Let $D$ be the sum of the remaining rational points. Then the code $C_\calL(D,G)$ is a $[63,39,25]$ two-point Reed-Solomon code over $\F_{64}$. We have $\dLO=25$.

Note that $n=63$, $g=0$ and $\gam=1$. The maximum degree of the polynomials in the vector forms of $h_i$ is $62=N_h$. The degree of the polynomial in the vector form of $\eta_0$ is $63=N_\eta$. In an experiment with $10^5$ instances of decoding random errors of weight $12$, the decoder performed at most $63=N_{\mathrm{iter}}$ iterations with $2\times 2$ matrix of univariate polynomials at most $63=N_{\mathrm{deg}}$ degree over $\F_{64}$. It took 0.0039 second to decode one instance, taking $O(3969)$ multiplications.

\section{Remarks}

We presented a unique decoding algorithm that can decode errors up to half of the bound $\dLO$. Beelen and H\o holdt's algorithm in \cite{beelen2007} is similar in approach to ours, and can decode up to half of their generalized order bound. Thus we can speculate that $\dLO$ is related with the generalized order bound. Indeed it was shown in \cite{geil2012b} that the bound $\dLO$ as defined in \cite{kwankyu2011} coincides with the so-called Andersen-Geil bound $d_{\mathrm{AG}}$ \cite{andersen2008}. The relationship between these bounds may be treated in a separate place.

Geil and et al. \cite{geil2012b} also showed that by a slight modification, the algorithm in \cite{kwankyu2011} can be turned to a list decoding algorithm. The same can be done with the present general algorithm, but we leave out the details.



\end{document}